\documentclass[preprint,onecolumn,nofootinbib]{revtex4}
\usepackage{graphicx}
\pdfoutput=1
\usepackage[colorlinks=true,linkcolor=blue,urlcolor=blue,filecolor=black,citecolor=red,pdfstartview=FitV,pdftitle={},pdfsubject={},pdfkeywords={},pdfpagemode=None,bookmarksopen=true]{hyperref}
\usepackage{epsfig}
\usepackage{amsmath}
\usepackage{amsfonts}
\usepackage{amssymb}
\usepackage{color}%
\usepackage{dcolumn}
\providecommand{\U}[1]{\protect\rule{.1in}{.1in}}

\newcommand{\f}{\begin{equation}}
\newcommand{\ff}{\end{equation}}
\newcommand{\fa}{\begin{eqnarray}}
\newcommand{\ffa}{\end{eqnarray}}

\begin{document}
\title{Building a doped Mott system by holography}
\author{Yi Ling $^{1,2}$}
\email{lingy@ihep.ac.cn}
\author{Peng Liu $^{1}$}
\email{liup51@ihep.ac.cn}
\author{Chao Niu $^{1}$}
\email{niuc@ihep.ac.cn}
\author{Jian-Pin Wu $^{3,2}$}
\email{jianpinwu@mail.bnu.edu.cn}
\affiliation{$^1$ Institute of High Energy Physics, Chinese Academy of Sciences, Beijing 100049, China\\
$^2$ State Key Laboratory of Theoretical Physics, Institute of Theoretical Physics, Chinese Academy of Sciences, Beijing 100190, China\\
$^3$Institute of Gravitation and Cosmology, Department of Physics,
School of Mathematics and Physics, Bohai University, Jinzhou 121013, China}
\begin{abstract}
We construct a holographic model in the framework of Q-lattices
whose dual exhibits metal-insulator transitions. By introducing an
interacting term between the Q-lattice and the electromagnetic
field in bulk geometry, we find such kind of transition can be
Mott-like. The evidences are presented as follows. i) The
transition from a metallic phase to an insulating phase occurs
when the lattice constant becomes larger. ii) A hard gap in the
insulating phase can be manifestly observed in the optical
conductivity. Nevertheless, in the zero temperature limit this
model exhibits novel metallic behavior, featured by a gap as well
as a zero-frequency mode with tiny spectral weight. It implies
that our model is dual to a doped Mott system in one dimension
where umklapp scattering is frozen at zero temperature. The
similarity between this model and some organic linear chain
conductors is briefly discussed.
\end{abstract}
\maketitle

\section{Introduction}

Gauge/gravity duality has provided powerful tools for
investigating strongly correlated systems in condensed matter
physics. In particular, the metal-insulator transition has
recently been implemented in this approach, in which the
translational symmetry is broken by various lattice
structures  \cite{DonosHartnoll,Donos:2013eha,Donos:2014uba,Ling:2014saa,Donos:2014oha,Baggioli:2014roa}.
However, until now it is still challenging to construct a
holographic model describing Mott insulator, which plays a crucial
role in a strongly correlated electron system and is always
responsible for the building of high $T_c$ cuprate superconductors
  \cite{XGWen:2006}. In this paper we intend to construct such a
holographic model whose dual can be viewed as a doped Mott system.

The notion of a Mott insulator was originally proposed by Mott in the 
1930s  \cite{Mott:1930s}. In his thought experiment, a metal could
become an insulator when increasing the lattice constant, simply
because the ability of electrons hopping from one site to its neighbors decays.
Mott's idea has been confirmed in
experiments (for instance, see
  \cite{Imada:1998,Limelette:2003,Drichko:2006} and references
therein.), in which the metal-insulator transition occurs when
changing external parameters such as the pressure which leads to
the change of lattice spacing. In the theoretical aspect, the Mott
thought experiment can be formalized in the Hubbard model
\cite{Hubbard:1963}. The Hamiltonian of Hubbard model contains two
key ingredients. One is the kinetic energy which describes the
electron hopping process and the other one is the potential energy
which describes on-site Coulomb interaction between two electrons
with opposite spin. Roughly speaking, the electron hopping process
is responsible for the metallic behavior while the repulsive
Coulomb interaction is for the insulating state. When the kinetic
and potential energies are of the same order, it is extremely
difficult to solve the Hubbard model exactly except for the
one-dimensional case. A powerful technique for solving the Hubbard
model is the dynamical mean field theory (DMFT)
\cite{DMFT1,DMFT2}, in which the lattice model is replaced by an
effective single impurity model.

In the holographic approach, dynamically generating a Mott gap has
been proposed in  \cite{Edalati:PRL,Edalati:2010ge,Ling:2014bda}
by considering fermions with dipole coupling over an
AdS-Reissner-Nordstr\"om (AdS-RN) geometry\footnote{It
is worthwhile to notice that the dual boundary theory of a
holographic system with dipole coupling fermions may suffer from
superluminal modes in large momentum limit \cite{Kulaxizi:2015fza}
.}. However, such a Mott gap is interaction driven and achieved
only in the probe limit. It should be stressed that the bulk
geometry of the background is completely dual to a metallic phase
because of the IR fixed point
being AdS$_2$. Later, novel metal-insulator transitions have been
implemented in black hole backgrounds with lattice structure. The
key point in this framework is to make the near horizon geometry
unstable such that the IR fixed point can deform from
AdS$_2$ to other saddle points which may be dual to an insulating
phase \cite{DonosHartnoll,Donos:2013eha,Donos:2014uba}. In this
paper we intend to push this approach forward and explicitly
construct a gravitational dual model which exhibits the
fundamental features of a one-dimensional doped Mott system.
Specifically, we will introduce an interacting term between the
Q-lattice and the electromagnetic field in bulk geometry. It turns
out that the transition from a metallic phase to an insulating
phase occurs when the lattice constant becomes larger, which
explicitly visualizes the thought experiment proposed by Mott in a
holographic manner. More importantly, the linear perturbation
analysis will demonstrate that a hard gap can be manifestly
observed in the optical conductivity\footnote{Near the completion
of this manuscript, we noticed the appearance of
\cite{Kiritsis:2015oxa} in which an insulator with a hard gap has
also been implemented in a holographic manner.}. Nevertheless, we
find this model always contains nonvanishing zero-frequency
modes. Although its spectral weight is a tiny proportion of the
total, it is responsible for the large metallic conductivity such
that the system exhibits a novel metallic behavior in zero
temperature limit. This behavior implies that our model is a doped
system where umklapp scattering is frozen at zero temperature.
Interestingly enough, we find our model is analogous to some
organic linear chain conductors which are featured by a gap as
well as nonvanishing Drude peak in metallic state
\cite{Dressel:1996,Vescoliet:1998}. We will briefly address this
issue in the end of our paper.

\section{The holographic setup}\label{Setup}
Our holographic setup is based on a Q-lattice structure, which has
originally been proposed in  \cite{Donos:2013eha} and subsequently
investigated in literature
  \cite{Donos:2014uba,Donos:2014oha,Ling14laa,Ling:2014bda,Ling:2015dma}.
Such a gravitational dual model contains a complex scalar field
and a $U(1)$ gauge field. We introduce the action as
\begin{eqnarray}
S&=&\frac{1}{2\kappa^2}\int
d^4x\sqrt{-g}[R+6-\frac{V(\Phi)}{4}F^{\mu\nu}F_{\mu\nu}-|\nabla
\Phi|^2-m^2|\Phi|^2], \label{action}
\end{eqnarray}
where we have fixed the $AdS$ length scale $L=1$. $\Phi$ is
uncharged and will be responsible for the breaking of the
translational symmetry. Comparing with the original action
proposed in  \cite{Donos:2013eha}, we have introduced a new
coupling $V(\Phi)$ between the scalar field and the gauge field,
which will play a crucial role in generating a Mott-gap in
insulating phase. For explicitness, in this paper we will set
$V(\Phi)=1-\beta |\Phi|^2$ with a positive parameter
$\beta$\footnote{We require that $\beta$ is positive such that the
system is always stable under linear perturbations. Otherwise a
negative $\beta$ would induce the instability of near horizon
geometry, leading to the spontaneous breaking of the translational
symmetry at low temperatures, as discussed in  \cite{Donos:2013gda,Ling:2014saa}.}.
Such a term is similar to that in  \cite{Mefford:2014gia}.
Obviously, the above construction goes back to the original
form in  \cite{Donos:2013eha} if one sets $\beta=0$. Therefore,
next we will skip all the technical steps for constructing a
Q-lattice background but just present the
results.

The ansatz for a background which has a Q-lattice structure only
along $x$ direction is given by \fa ds^2&=&{1\over
z^2}\left[-(1-z)p(z)Udt^2+\frac{dz^2}{(1-z)p(z)U}+V_1dx^2+V_2dy^2\right],\nonumber\\
A&=&\mu(1-z)\psi dt,\nonumber\\
\Phi &=& e^{ikx}z^{3-\Delta}\phi,\ffa where $\mu$ is the chemical
potential, $p(z)=1+z+z^2-\mu^2z^3/4$ and
$\Delta=3/2+(9/4+m^2)^{1/2}$. All the functions $U,V_1,V_2,\psi$
and $\phi$ depend on the radial coordinate $z$ only. Through this
paper we set $m^2=-2$. For a given $\beta$, each electrically
charged black hole solution with Q-lattice is specified by three
scaling-invariant parameters, namely, the Hawking temperature
$T/\mu$ which is given by  $(12-\mu^2)U(1)/(16\pi\mu)$, the
lattice amplitude $\lambda/\mu^{3-\Delta}$ where
$\lambda\equiv\phi(0)$ and the wave vector $k/\mu$. For
convenience, we abbreviate these quantities to $T$, $\lambda$
and $k$, respectively. Moreover, it is worthy to point out that
in this formalism the background has a periodic structure
manifestly such that $a\equiv 2\pi/k$ can be understood as the
lattice constant of the dual model, which is important for us to
link this holographic scenario with the thought experiment by Mott
in next section.

\section{Phase diagram and optical conductivity}

It is first shown in \cite{Donos:2013eha} that the dual system
of a Q-lattice background exhibits metal-insulator
transitions when varying the lattice parameters $\lambda$ and $k$,
which corresponds to the special case  with $\beta=0$ in our
model. When the parameter $\beta$ is turned on we find that the
coupling term will play the role of generating a hard gap for the
optical conductivity, which is analogous to the effect of Coulomb
interaction $U$ in Mott-Hubbard model. We intend to discuss the
phase diagram and the optical properties of the system in
following subsections.

\subsection{Finite temperature region}

In this subsection we investigate the $(\beta,k)$ phase diagram
and the optical properties of the dual system when the temperature is
not too low. We will demonstrate that a metal-insulator transition
occurs when the lattice constant becomes larger. Furthermore, a
Mott hard gap can be manifestly observed in the plot of the
optical conductivity in insulating phase when the parameter
$\beta$ is adjusted to an appropriate value.

\begin{figure}
\center{
\includegraphics[scale=0.6]{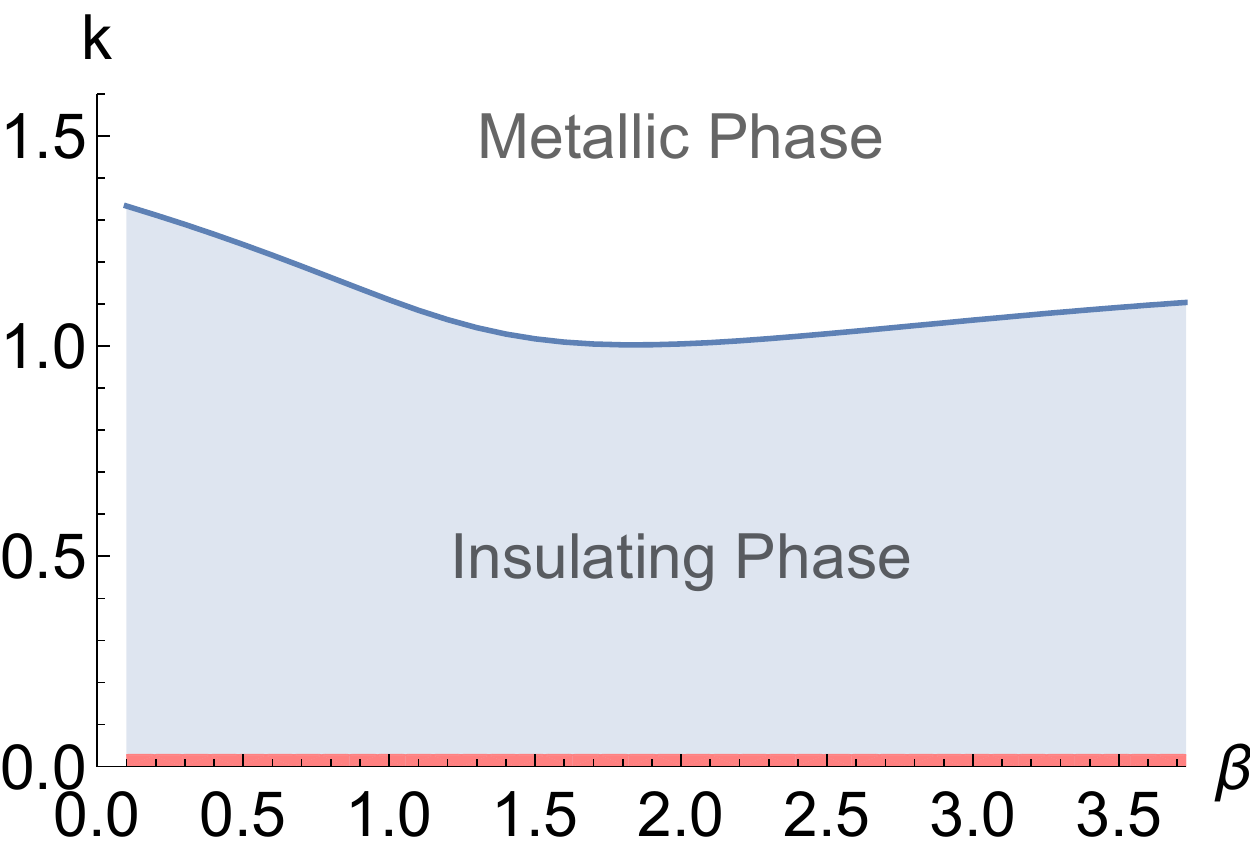}\\ \hspace{0.4cm}
\caption{\label{phasekb2}The phase diagram with ($\lambda=2,\; T=0.2$). Our numerical data are collected beyond the pink region with $k<0.03$.}}
\end{figure}

First, we plot the ($\beta$,$k$) phase diagram at the
temperature $T=0.2$, as illustrated in Fig.\ref{phasekb2}. The
insulating phase corresponds to $\sigma_{DC}'(T)>0$ while
the metallic phase $\sigma_{DC}'(T)<0$. Thus the critical
line is described by $\sigma_{DC}'(T)=0$. $DC$ conductivity can
be obtained by either the zero-frequency limit of optical $AC$
conductivity, or the numerical analysis on the horizon with
formula ${\sigma _{DC}} = \sqrt {\frac{{{V_2}}}{{{V_1}}}} {\left.
{\left[ {V\left( \Phi  \right) + \frac{{{V_1}}}{2}{{\left(
{\frac{{V\left( \Phi  \right)\mu \psi }}{{k\phi }}} \right)}^2}}
\right]} \right|_{z = 1}}$. In this phase diagram, one can easily
find that a metal-insulator transition occurs when the wave number
deceases (or the lattice constant increases) for a given lattice
amplitude $\lambda$ and coupling parameter $\beta$, which could be
viewed as a holographic realization of Mott thought experiment.
Therefore, the $(\beta,k)$ phase diagram provides us the first
evidence that the resulting metal-insulator transition in our
model is Mott-like, which is induced by changing the lattice
constant.

Next we provide the second key evidence for the Mottness of this
sort of insulators by a linear perturbation consideration. We
compute the optical conductivity as a function of frequency
when applying an external electric field along $x$ direction. In
holographic setup this can be achieved by introducing the
following self-consistent perturbations over the background,
\[\delta g_{tx}=h_{tx}(t,z),\quad \delta A_x = a_x (t,z),\quad \delta\Phi = i e^{ikx} z^{3-\Delta}\phi(t,z),\]
where the time dependence are all set as $e^{-i\omega t}$, such
that the optical conductivity is given by
$\sigma(\omega)=\left.\frac{\partial_z
a_x(z)}{i\omega a_x (z)}\right|_{z=0}$.

\begin{figure}
\center{
\includegraphics[scale=0.5]{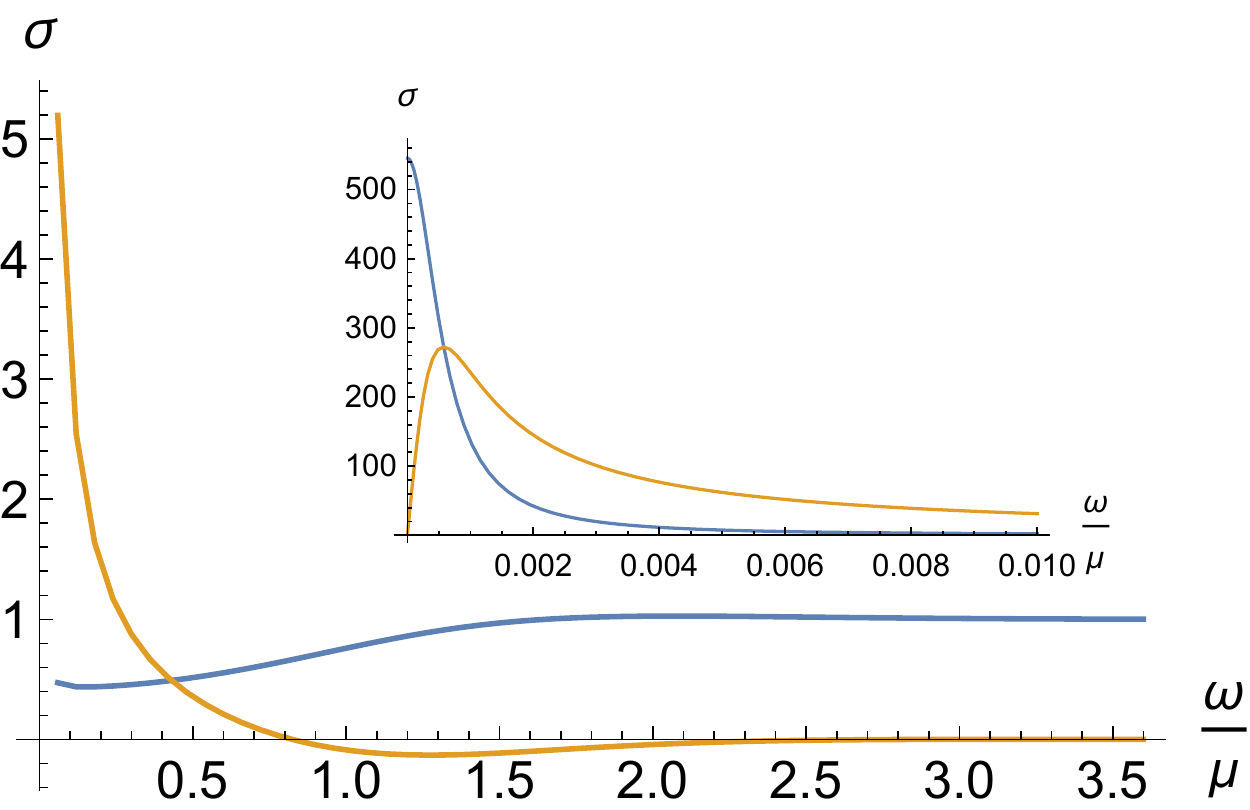} \hspace{0.4cm}
\includegraphics[scale=0.5]{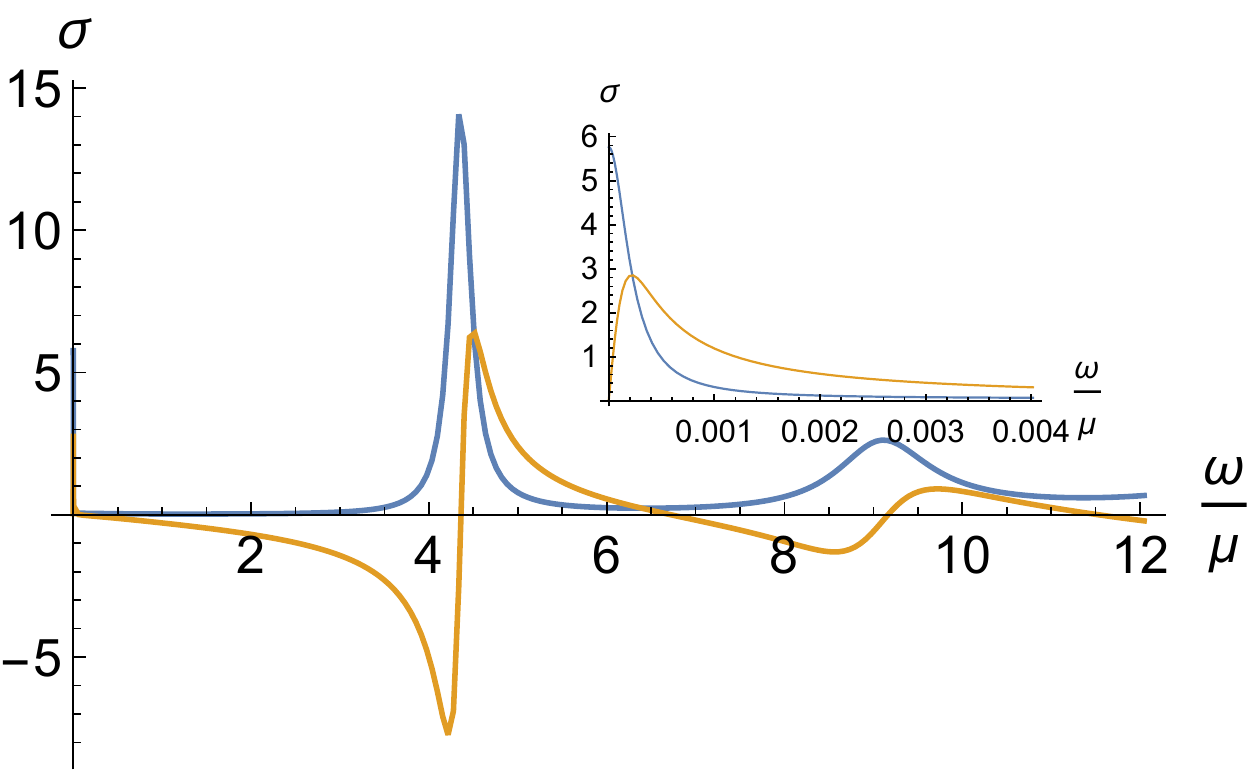} \hspace{0.4cm}
\caption{\label{op_T0p05} The real part (blue curves) and the
imaginary part (brown curves) of optical conductivity with
different values of $\beta$ (the left plot is for $\beta=0$ while
the right one for $\beta=3$).  The other parameters are fixed as
$\lambda=2$, $k=0.03$ and $T=0.2$. The insets in both plots
are the blow-up of the optical conductivity in low
frequency region.}}
\end{figure}

We explicitly demonstrate that a hard gap can be produced in
insulating phases with the increase of the parameter $\beta$, as
illustrated in Fig.\ref{op_T0p05}. The left plot in
Fig.\ref{op_T0p05} is for $\beta=0$, only a soft gap being
observed. Previously, a similar phenomenon has also been presented
in  \cite{DonosHartnoll,Donos:2013eha}. While the right plot of
Fig.\ref{op_T0p05} is for $\beta=3$, in which a hard gap can be
obviously seen. It is one of the important characteristics of a
class of Mott insulators. In particular, it is usually
believed that in the Mott-Hubbard model, the formation of a hard
gap is due to the localization of electrons driven by the Coulomb
interaction $U$. In previous holographic insulator models,
only a soft gap is observed in the optical conductivity
\cite{DonosHartnoll,Donos:2013eha,Donos:2014uba}. Here we provide
a new localization mechanism for the gap formation by holography.
In this mechanism, the coupling parameter $\beta$ plays a crucial
role, which is similar to the Coulomb interaction $U$.
In addition, we would like to present the following remarks on the
optical conductivity when we change the value of the coupling
parameter $\beta$.
\begin{itemize}
    \item {} The gap becomes more evident with the increase of $\beta$.
   This phenomenon is also in accordance with the effect of
    increasing $U$ in Mott-Hubbard model since the enhancing
    strength of electron-electron interaction will localize electrons.
    If one extracts the charge density $\rho$ from the time component of the
    gauge field in background solutions, it is also found that the charge density decreases
    with the increase of $\beta$, which possibly provides an understanding
    on the suppression of the optical conductivity and the emergence of a
    hard gap.

\item {} The insets in Fig.\ref{op_T0p05} illustrate the low
frequency behavior of the optical conductivity.
    It turns out that generically the low frequency behavior can be fit well with the Drude
    formula. This phenomenon asks for further understanding since
    now the dual system lies in the insulating phase. Moreover, comparing two insets in Fig.\ref{op_T0p05}
    we obviously see that the Drude weight diminishes with the increase of $\beta$.

\item {} With the increase of $\beta$, more resonance peaks show
up in the intermediate frequency region of the optical
conductivity and approximately exhibit a periodic distribution (as
is seen in the right plot in Fig.\ref{op_T0p05}), indicating a
multiband structure for the dual theory.

\end{itemize}

Next, we study the properties of our model in zero temperature limit.

\subsection{In zero temperature limit}

\begin{figure}
\center{
\includegraphics[scale=0.6]{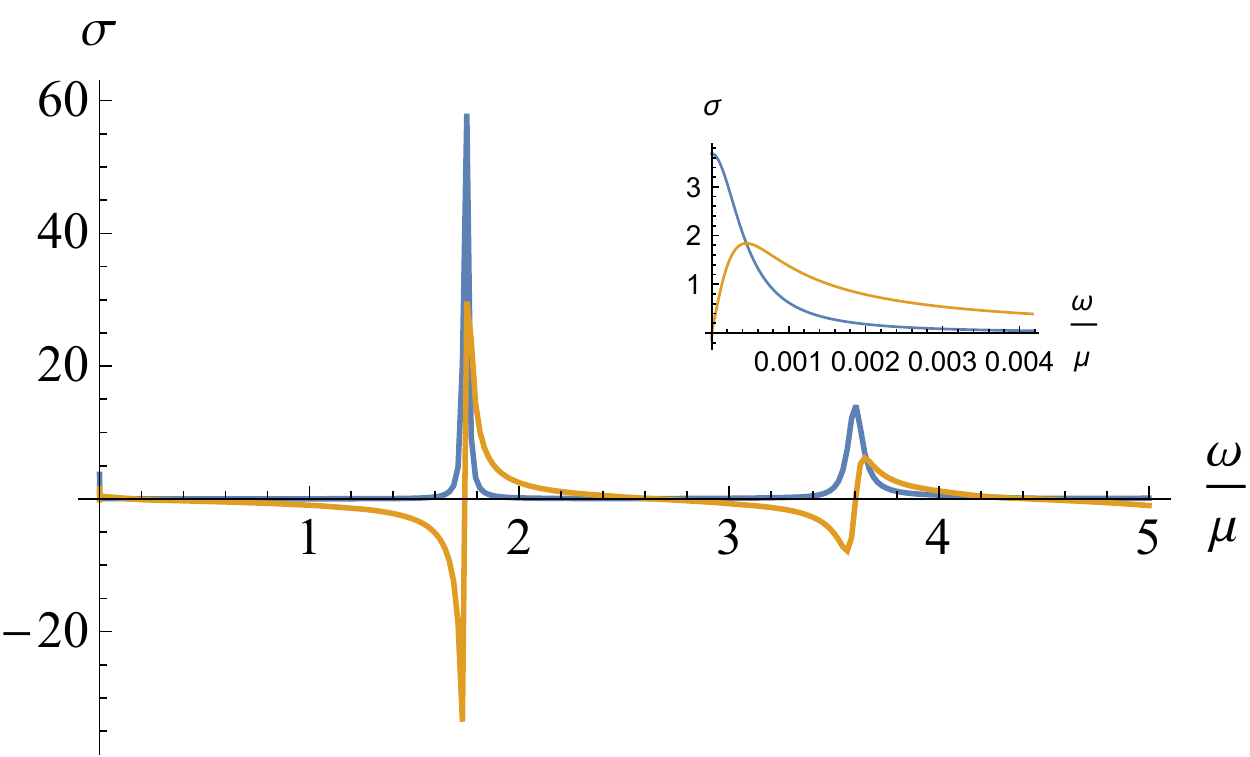} \hspace{0.4cm}
\caption{\label{op_T0p004} The real part (blue curves) and the
imaginary part (brown curves) of optical conductivity for
$\lambda=2$, $k=0.03$, $\beta=0.766$ and $T=0.004$. The inset
is the low frequency behavior of the conductivity.}}
\end{figure}

To explore the characteristics of this system thoroughly, we
cool it to extremely low temperature. As an example, we show the
optical conductivity at $T=0.004$ in Fig.\ref{op_T0p004}, which is in an insulating phase.
It is noticed that the appearance of a hard gap as well as
resonances resembles the phenomena in the second plot of
Fig.\ref{op_T0p05} except that the resonance peaks become sharper.
Moreover, the low frequency behavior of the optical conductivity
(shown in the inset) can be well fit with Drude formula as well.

\begin{figure}
\center{
\includegraphics[scale=0.6]{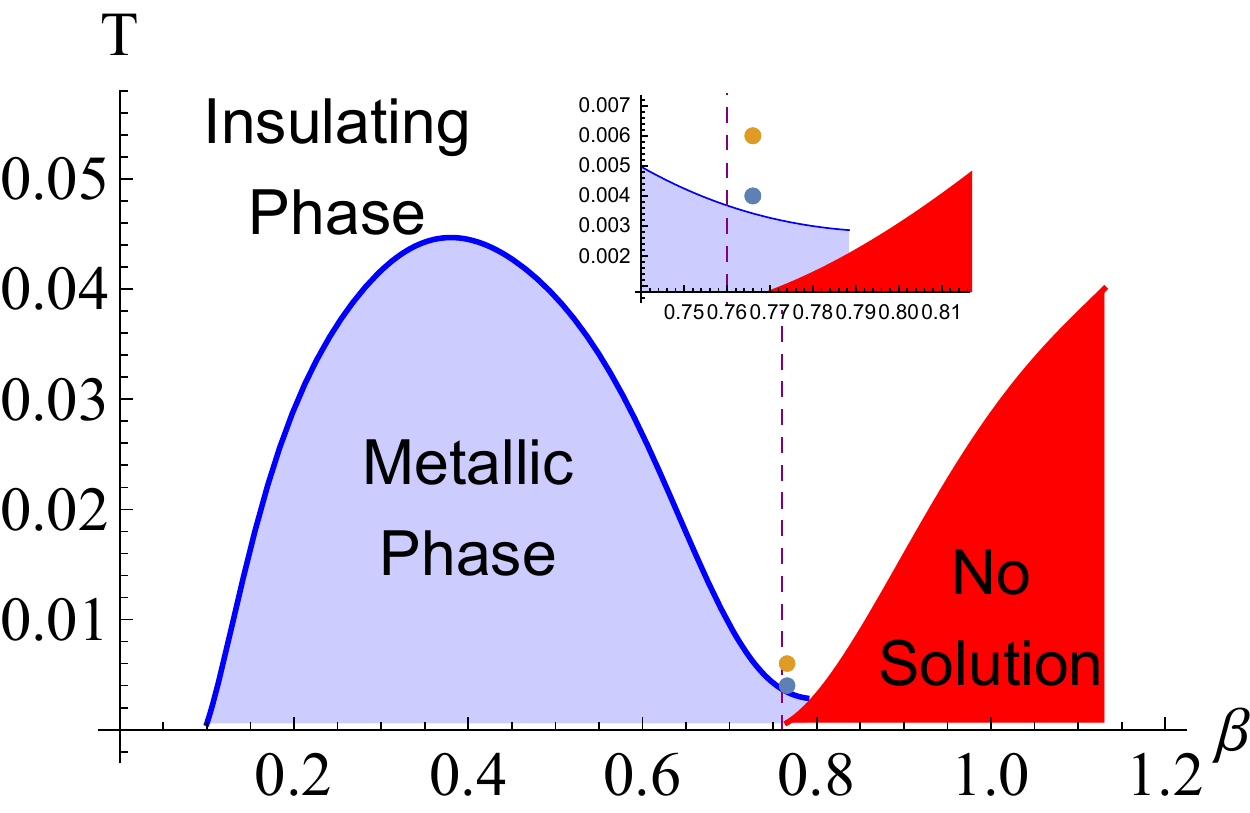}\hspace{0.5cm}
\caption{\label{Phase}$(\beta, T)$ phase diagram with $\lambda=2,
k=0.03$. The inset is a blow up of the intersecting region nearby
$\beta=0.78,T=0.003$. The purple dashed line and two dots
denote the parameters adopted in Fig.\ref{rhot}.}}
\end{figure}

\begin{figure}
\center{
\includegraphics[scale=0.6]{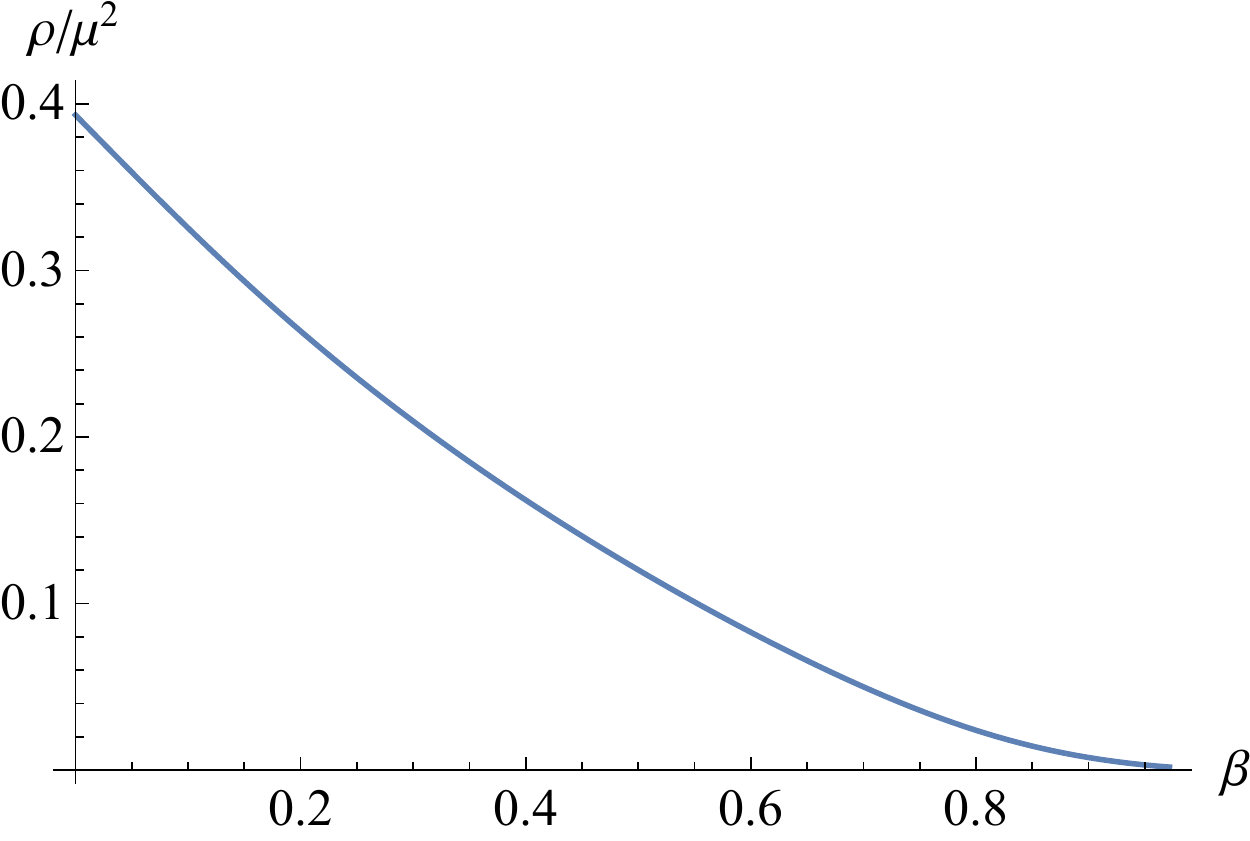}\hspace{0.5cm}
\caption{\label{rhobeta} Charge density $\rho/\mu^2$ as
a function of $\beta$ at $k=0.03,\lambda=2,T=0.03$.}}
\end{figure}

\begin{figure}
\center{
\includegraphics[scale=0.6]{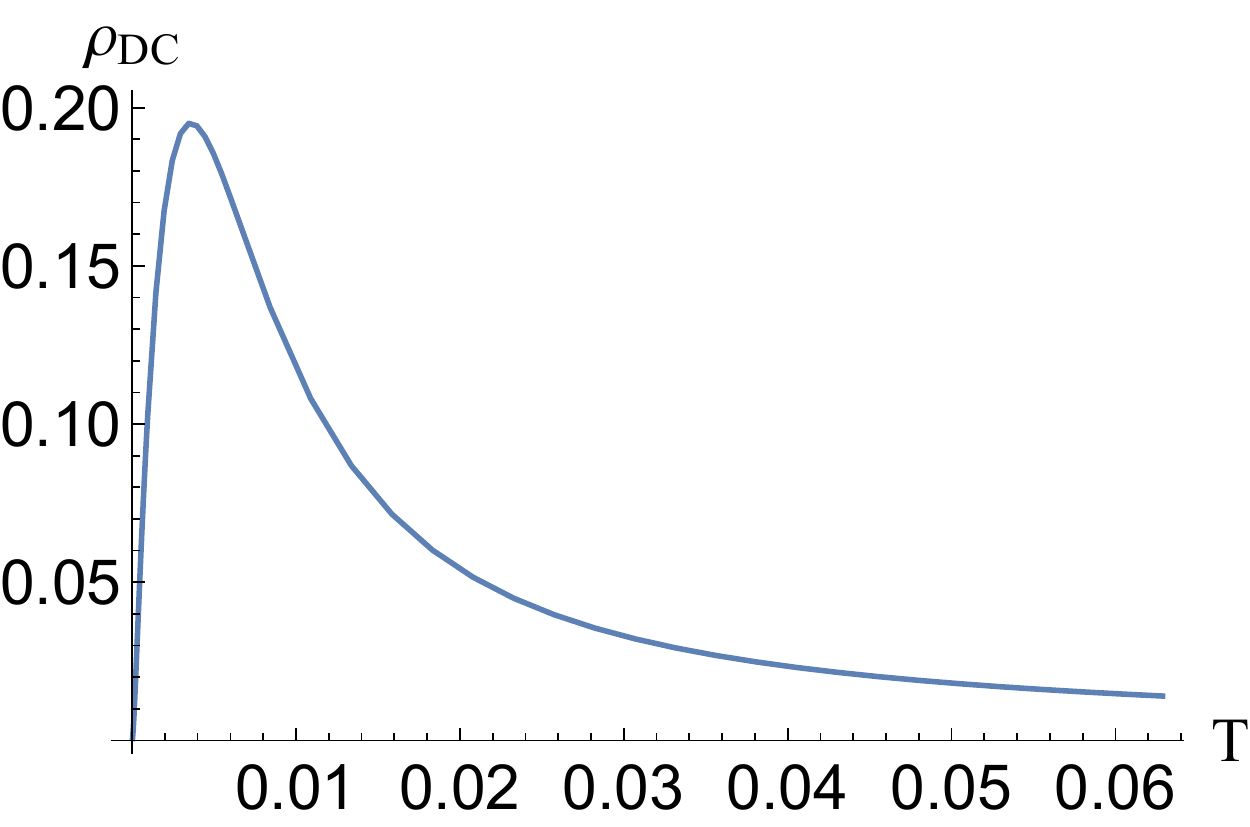} \hspace{0.4cm}
\includegraphics[scale=0.8]{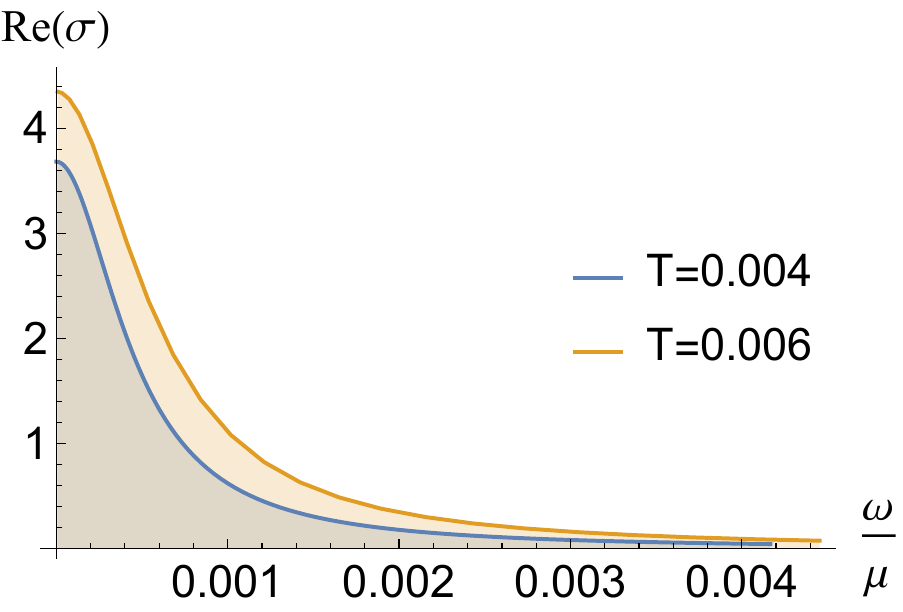} \hspace{0.4cm}
\caption{\label{rhot} The left plot is for the temperature
dependence of DC resistivity $\rho_{\text{DC}}$ at extremely low temperature region
for the selected $\lambda=2$, $k=0.03$, $\beta=0.76$ (which
corresponds to the purple dashed line in Fig.\ref{Phase}.). The
right plot shows the real part of optical conductivity at
temperatures $T=0.004 \text{ and } 0.006$, with $\lambda=2$,
$k=0.03$, $\beta=0.766$ (which are marked as the dots in the
inset of Fig.\ref{Phase}).}}
\end{figure}

In order to explore the quantum critical behavior of the
system, we intend to study the structure of phase diagram in zero
temperature limit. For definiteness, we plot the $(\beta,T)$
phase diagram with $\lambda=2$ and
$k=0.03$, as shown in Fig.\ref{Phase}. We present our remarks on
this phase diagram in order as follows.
\begin{itemize}
\item The phase diagram is divided into three regions
corresponding to the metallic phase, insulating phase and
no-solution region, respectively. The critical line between the
metallic phase and the insulating phase is determined by
$\sigma_{DC}'(T)=0$. Numerically, we find the charge density
$\rho$ is always decreasing with the increase of $\beta$, and we
show an example in Fig.\ref{rhobeta}. As $\rho$ becomes vanishing,
no numerical solution to background equations could be found when
further increasing the parameter $\beta$, therefore a no-solution
region forms as shown in Fig.\ref{Phase}. The interface between
the no-solution region and the other two regions can roughly be
described by $\rho\simeq 0$. From the viewpoint of dual field
theory, the fact of $\rho\to 0$ might correspond to $U\to \infty$
in Mott-Hubbard model since in this limit all the electrons will
be localized. Moreover, we point out that the coupling term
$V(\Phi)$ should be positive everywhere to avoid ghosty gauge
fluctuations. We have checked the behavior of $V(\Phi)$ over all
the $z$-axis and found that in usual cases it is always positive.
When $\beta$ becomes large and particularly points to the
borderline of no-solution region, $V(\phi)$ approaches to but
never below $0$ in a small region of $z$. To guarantee our model
is safe we have also checked the case for $V(\Phi)=(1-\beta
|\Phi|^2)^2$ which is positive definite, and we qualitatively
recovered all the results and phenomena in the current paper.

\item Numerically we have found that the hard gap is not evident
until the parameter $\beta$ is adjusted to a relatively large
value $\beta_c$. In Fig.\ref{Phase}, for parameters $\lambda=2$
and $k=0.03$ we find this critical value is about $\beta_c\simeq
0.4$.

\item{} We are very concerned with the intersecting part of these
three regions in zero temperature limit because it provides us a
hint on whether the system could fall into an insulating region
with hard gap at zero temperature. Therefore, we blow up the
intersecting region as shown in the inset of Fig.\ref{Phase}.
Unfortunately from this inset we notice that there is no such an
insulating region in the zero temperature limit.

\item To demonstrate the metal-insulator transition in a more
transparent manner, we plot the DC resistivity $\rho_{\text{DC}}$ as a function of
temperature in the left plot of Fig.\ref{rhot}, with parameters
specified by the purple line in Fig.\ref{Phase}. It is observed
that the $\rho_{\text{DC}}$ rises up with the decreasing temperature
at first, but after reaching its maximum it turns to go down and
eventually falls into zero when approaching the zero temperature.
Interesting enough, this behavior is much similar to the one as
described in  \cite{Georges:2004} when the system is close to the
Mott transition.

\item The low frequency behavior of the optical conductivity can
always be fit well with the Drude formula whenever a hard gap is
evidently observed. However, the Drude weight is always
diminishing as the temperature decreases. For explicitness, we
show the low frequency region of the real part of conductivity at
temperatures $T=0.006$ and $T=0.004$ in the right plot of
Fig.\ref{rhot}. In this figure the decrease of the Drude
weight with the temperature is manifest. We find this tendency is
also true for a metallic phase. That is to say, although the DC
conductivity becomes larger with the decrease of the temperature
in metallic phases, the Drude weight is always diminishing.
Nevertheless, the Drude weight does not vanish in the
zero-temperature limit, whatever the phase is. This phenomena
could be reflected as the feature of a doped Mott system or a
deviation from the commensurate filling in one dimension as we
will elaborate below.

\item Finally, we intend to claim that all above observations do
not change qualitatively with other choices of parameters
$(\lambda,k)$. For instance, we may have a larger value of
$\lambda$, then the metallic region may shrink into a smaller
region and the critical value $\beta_c$ for an evident hard
gap could be smaller, while at the same time one would find the
no-solution region expands such that the insulating phase
with hard gap remains absent in zero temperature limit.
\end{itemize}


In the end of this section we intend to focus on an
important issue related to understanding the emergence of the metallic
phase region in zero temperature limit, as illustrated in
Fig.\ref{Phase}. Theoretically, the presence of a gap as well as
a tiny spectral weight in low temperature limit is in accordance
with a doped Mott system as described in
\cite{Giamarchi:1997,Giamarchi:book}. First, in high frequency
region there are interband transitions and the doping will not
affect the system. Thus, the high frequency behavior will be the
same as the commensurate filling. However, for the frequencies
being smaller, the interband transitions are blocked, and then the
conductivity becomes vanishing, leading to the formation of a gap.
Second, when a system is doped, the Fermi energy may not fall
into the gap region such that intraband transitions are still
possible and result in nonzero low frequency excitations, which
is supposed to be responsible for the metallic behavior of the
system. Though the Drude weight is very tiny in the spectrum, its
nonvanishing feature could give rise to a divergent DC
conductivity in zero temperature limit. From a microscopic point
of view,  this phenomenon could be understood as a consequence of
doping which deviates the system from commensurate filling such
that the umklapp scattering is frozen in a one dimensional system.

The gapped behavior as well as a tiny Drude weight of optical
conductivity have been observed in some organic materials such as
the TMTSF family  \cite{Dressel:1996,Vescoliet:1998}, which is
analogous to a doped Mott system  \cite{Giamarchi:1997}. The
organic conductors are quasi-one dimensional systems with a
perpendicular hopping between chains. It is also argued in
  \cite{Giamarchi:1997} that doping could be attributed to the
transverse hopping between chains, which results in deviation from
commensurate filling because of the warping of the Fermi surface.
Nevertheless, we need to point out that for such organic
conductors the metallic behavior of the zero frequency modes due
to the interchain transitions are always decoherent and deviated
from a Drude law  \cite{Dressel:1996}, which seems in contrast to
what we observed here.

\section{Discussion}

In this paper we have constructed a holographic model based on
Q-lattice which can be dual to a doped Mott system. The key
ingredient in our proposal is the introduction of a coupling term
between the Maxwell field and the scalar field in the Q-lattice
geometry. We have explicitly demonstrated that the Mott thought
experiment can be visualized by holography in Q-lattice framework.
More importantly, a hard gap can be produced in the optical
conductivity when the coupling parameter $\beta$ is relatively
large. While in the zero temperature limit we find the Drude
weight is not vanishing and the system exhibits a novel metallic
behavior. All above features are closely similar with those of a
doped Mott system in one dimension, and have been observed in the
TMTSF family of organic materials.

Our results reported here are preliminary and should be thought of
as the first step for building a holographic model for Mott
insulator which is usually commensurate. To gain more insight 
into this topic, a lot of work needs to be done. Especially, the
following crucial issues should be explored further. First, it
is very desirable for one to understand why the coupling
term plays the role of generating a hard gap for the optical
conductivity, which is very crucial for us to figure out the
direct connection between the coupling term in this holographic
model and the electron-electron interaction rooted in a realistic
Mott system. We expect an analytical investigation on the
current-current correlator and the deformation of the near horizon
geometry like
\cite{Charmousis:2010zz,Gouteraux:2014hca,Davison:2015bea}, would
be helpful for us to understand how the coupling term changes the
gapless geometry to gapped one, as well as the multiband
structure and the emergence of novel metallic phase in the zero
temperature limit. Alternatively, one could provide more evidences
for doping if the spectral weight transfer could be found in the
fermionic spectral function including an appropriate coupling
between fermions and the lattice background. Second, we are
crying for some new  mechanism in Q-lattice framework to suppress
the Drude weight so as to achieve an insulating phase with a hard
gap in the zero temperature limit. Finally we remark that
throughout this paper we have only considered Q-lattice structure
in one spatial direction. It is very worthwhile to investigate the
Mottness of the dual system in two or higher dimensional Q-lattice
background or other lattice models \cite{Rangamani:2015hka}. Our
work in this direction is under progress.

\begin{acknowledgments}
We are grateful to Matteo Baggioli, Aristomenis Donos, Jerome
Gauntlett, Blaise Gout\'eraux, Sean Hartnoll, Honggang Luo, Philip
Phillips, Jie Ren, Yu Tian, Yidun Wan, Yuan Wan, Zhuoyu Xian,
Yifeng Yang, Hongbao Zhang for helpful discussions. We also
thank the anonymous referee for detailed comments and helpful
suggestions. This work is supported by the Natural Science
Foundation of China under Grants No. 11275208, No. 11305018 and
No. 11178002. Y.L. also acknowledges the support from Jiangxi young
scientists (JingGang Star) program and 555 talent project of
Jiangxi Province. J. P. Wu is also supported by the Program for
Liaoning Excellent Talents in University (No. LJQ2014123). This
research was supported in part by Perimeter Institute for
Theoretical Physics. Research at Perimeter Institute is supported
by the Government of Canada through Industry Canada and by the
Province of Ontario through the Ministry of Economic Development
and Innovation.

\end{acknowledgments}

\end{document}